\documentclass[onecolumn,prb]{revtex4}
\usepackage{graphicx,psfrag,dsfont}
\usepackage{slashed,amssymb}
\def\no{\noindent}
\def\bc{\begin{center}}
\def\ec{\end{center}}

\def\beq{\begin{equation}}
\def\eeq{\end{equation}}

\def\d{\downarrow}
\def\u{\uparrow}

\def\br{{\bf r}}
\def\bq{{\bf q}}
\def\bk{{\bf k}}
\def\bp{{\bf p}}

\def\bR{{\bf R}}

\begin{document}

\title{
Pairing transition in a double layer with interlayer Coulomb repulsion
}
\author{Andreas Sinner$^{1,2}$, Yurii E. Lozovik$^{3,4}$, and Klaus Ziegler$^{1}$}
\affiliation{
\mbox{$^{1}$ Institut f\"ur Physik, Universit\"at Augsburg, D-86135 Augsburg, Germany}\\
\mbox{$^{2}$ Max-Planck-Institut f\"ur Physik komplexer Systeme, D-01187 Dresden, Germany}\\
\mbox{$^{3}$ Institute of Spectroscopy, Russian Academy of Sciences,142190 Troitsk, Moscow, Russia}\\
\mbox{$^{4}$ Moscow Institute of Electronics and Mathematics, National Research University}\\
\mbox{Higher School of Economics, 101000 Moscow, Russia}
}
\date{\today}

\begin{abstract}
We study the effect of interlayer Coulomb interaction in an electronic double layer.
Assuming that each of the layers consists of a bipartite lattice, a sufficiently strong
interlayer interaction leads to an interlayer pairing of electrons with a staggered
order parameter. We show that the correlated pairing state is dual to the excitonic
pairing state with uniform order parameter in an electron-hole double layer. The
interlayer pairing of electrons leads to strong current-current correlations between
the layers. We also analyze the interlayer conductivity and the fluctuations of the 
order parameter, which consists of a gapped and a gapless mode.
\end{abstract}

\maketitle

\no
Layered electronic systems have attracted substantial attention over several decades because new physical
effects can be observed, which neither exist in a single layer nor in an isotropic three-dimensional
systems. The role of the layered structures might be important for physical systems ranging from high-T$_c$ 
superconductors \cite{bednorz86} to new quantum phases in twisted bilayers with a ``magic angle'' \cite{cao18a,cao18b}. 
Another direction of recent research is associated with multilayer graphene \cite{murata19} and transition metal dichalcogenide 
(TMDC) multilayers \cite{geim13}, where an anomalous giant magnetoresistance \cite{song18}, superconductivity \cite{saito16}
and the formation of exciton condensates \cite{butov01,kellogg02,spielman04,tutuc04,qiu13,fogler14} have been discussed and observed. 
Other interesting effects in layered systems are based on the application of an external magnetic field. In recent experiments 
the emergence of novel correlated many-body states between 
quasiparticles from different layers was observed~\cite{Dean2017,Halperin2019}. 
In this Letter we propose electron interlayer pairing caused by a repulsive interlayer interaction. As we will 
show in a mean-field calculation, there is a second order phase transition from two uncorrelated Fermi liquids 
in the two layers to a correlated pairing state if a critical interaction strength is exceeded. Moreover, we discuss
a duality transformation between the pairing states of an electronic double layer and the excitonic pairing
state in an electron-hole double layer. 

\vspace{1mm}
\no 
\section{ Model} 
In an isolated (e.g., graphene-like) two-dimensional layer the electrons are subject to hopping when we assume that the 
Coulomb interaction within the layer is screened and renormalizes only the hopping parameters~\cite{Vozmediano1994,Vozmediano1999,Vozmediano2012}.
In the case of two parallel layers there is also a Coulomb interaction which acts between the electrons of the two layers.
This interlayer Coulomb interaction can be adjusted by inserting a dielectric material between the layers. Interlayer
tunneling is ignored here to demonstrate only the effect of the interlayer Coulomb interaction. It may have some effect on
the form of the order parameter though, as it was observed in the case of an attractive interlayer Coulomb interaction \cite{kettemann92}.
This model has some similarity with exciton models in double layers \cite{berman19}, where we have electrons in one and holes 
in the other layer. While the Coulomb interaction between electrons and holes is attractive, the Coulomb interaction is repulsive 
in the electronic case. This implies that we do not expect the formation of excitonic Cooper pairs but a collective state build by electron 
pairs in the two layers, provided the Coulomb interaction is strong enough and the thermal fluctuations are weak at sufficiently 
low temperatures. Then the electronic Hamiltonian consists of the two independent hopping terms $H_\u$ (upper layer) and $H_\d$ 
(lower layer) and the repulsive interaction of the electrons between the two layers $H_I$ with
\begin{equation}
H^{}_I = \sum_{\br,\br^\prime} n^{}_{\uparrow\br} U^{}_{\br,\br^\prime} n^{}_{\downarrow\br^\prime} +h.c.
\ , 
\label{interaction00}
\end{equation}
where $\br$ and $\br^\prime$ are lattice sites of each layer and $n_{s\br}$ is the local electron density operator 
in the layer $s=\u,\d$. The interlayer Coulomb interaction $U^{}_{\br,\br^\prime}$ is short-ranged in the lateral direction
due to screening \cite{kulakovskii04,lozovik76} and can be approximated by a diagonal matrix
$ U^{}_{\br,\br^\prime} \approx g \delta^{}_{\br,\br'}/2 $,
where $g$ is an effective parameter that decreases with the distance between the layers and depends on the 
interlayer dielectric material.
This approximation is valid for interlayer distance large in comparison with the lattice constant of the layers,
as described in Appendix~\ref{app:EffInt}, where also the dependence of $g$ on the various physical parameters is given.
The resulting total Hamiltonian $H=H_\u+H_\d+H_I$ has the structure of the Hubbard Hamiltonian, provided
that we interpret the layer index as the $z$-component of the electronic spin. 

For a bipartite lattice (e.g., a honeycomb lattice or simple square lattice) at half filling we introduce a sublattice representation 
of the coordinates $\br=(\bR,j)$ with $j=1,2$ for sublattice A and B, respectively. The coordinates $\bR$ refer to sublattice A only.
Then the hopping Hamiltonians $H_s$ ($s=\u,\d$) read in terms of fermionic creation (annihilation) operators $\hat c^\dagger_{s;\bR}$
($\hat c^{}_{s;\bR}$)
\begin{equation}
\label{eq:tbh00}
H^{}_s =  \sum^{2}_{j=1}\sum_{\bR,\bR'}h^{}_{j;\bR\bR'} \hat c^\dagger_{s;\bR}\cdot\sigma^{}_j\hat c^{}_{s;\bR'},
\end{equation}
where $\sigma^{}_{j=1,2}$ are Pauli matrices. The fermionic annihilation operators are written as column vectors
$\displaystyle
\hat c^{}_{s;\bR} =
\left(
\begin{array}{c}
 c^{}_{s;\bR} \\
 d^{}_{s;\bR}
\end{array}
\right),
$
whose upper (lower) component refers to sublattice A (B). The interaction term (\ref{interaction00}),
together with the diagonal approximation, 
then becomes
\begin{equation}
\label{interaction}
H^{}_I = \frac{g}{2}\sum_{\bR} n^{}_{\uparrow\bR} n^{}_{\downarrow\bR}
\ ,\ \ 
n^{}_{s\bR}=c^\dagger_{s\bR}c^{}_{s\bR}+d^\dagger_{s\bR}d^{}_{s\bR}
\ .
\end{equation}

\no
\subsection{Mean-field analysis} 
As a possible ansatz for the local order parameter we consider a staggered order parameter with opposite sign on
the two sublattices 
\begin{equation}
\Delta^{}_{\u\d;j}= -(-1)^j\Delta
\ , \ \ 
\Delta^{}_{\d\u;j}=\Delta^{*}_{\u\d;j}
\ ,
\end{equation}
where a global phase of $\Delta$ reflects the global $U(1)$ invariance of the model. A special mean-field solution fixes this phase.
For simplicity, we can choose a real positive $\Delta$ for the subsequent calculation.
The staggered order parameter is similar to the order parameter of an antiferromagnetic phase in the Hubbard model.
\begin{figure*}[t]
\includegraphics[width=7.5cm]{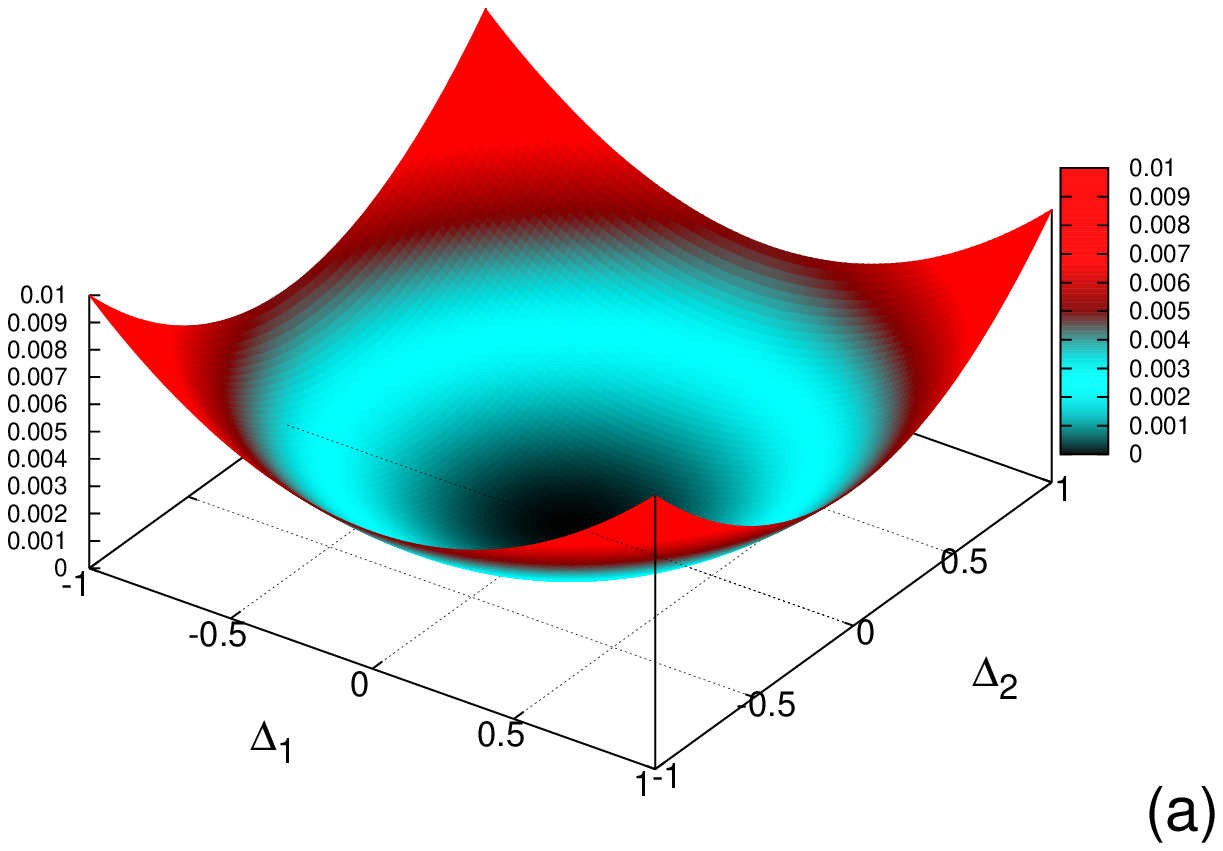}
\includegraphics[width=7.5cm]{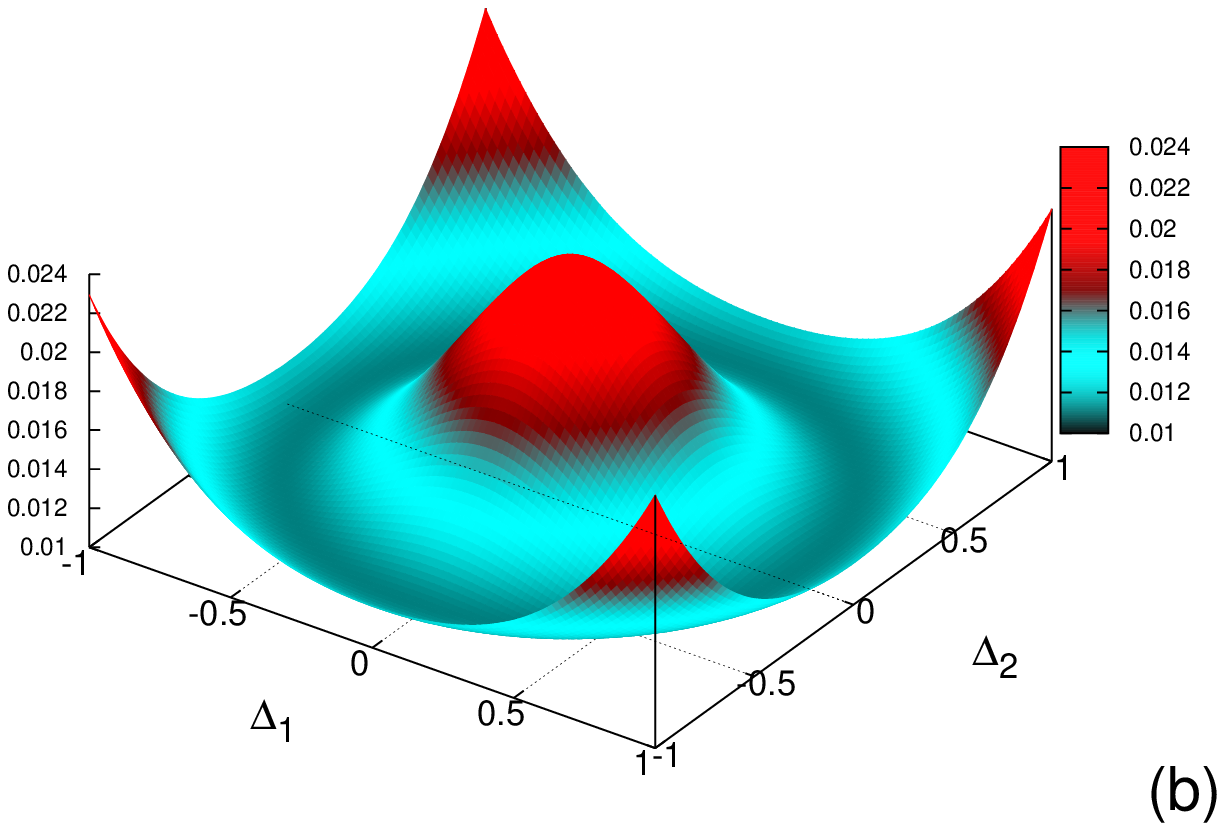}
\caption{
Mean-field potential. (a): Below critical interaction strength $g_c$. (b): Above critical interaction strength $g_c$.
}
\label{fig:Potential}
\end{figure*}
The staggering sign translates into a $\sigma_3$ Pauli matrix in the
corresponding quadratic mean-field Hamiltonian, cf. Appendix~\ref{app:MeanField}: 
\begin{equation}
\label{eq:MFH}
{\rm H}^{}_{\rm MF} = 
\sum_{\bR\bR'}
\left( 
\begin{array}{c}
\hat c^\dag_{\uparrow;\bR} \\ \\
\hat c^\dag_{\downarrow;\bR}
\end{array}
\right)^{\rm T}
\left(
\begin{array}{ccc}
\sum_j h^{}_{j;\bR\bR'}\sigma^{}_j & & \Delta \sigma^{}_3\delta^{}_{\bR\bR'} \\
 \\
 \Delta\sigma^{}_3\delta^{}_{\bR\bR'} & & \sum_j h^{}_{j;\bR\bR'}\sigma^{}_j
\end{array}
\right)
\left( 
\begin{array}{c}
\hat c^{}_{\uparrow;\bR'} \\ \\
\hat c^{}_{\downarrow;\bR'}
\end{array}
\right).
\end{equation}
In the staggered order parameter we can replace $(-1)^j$ by a general phase factor $\exp(i\varphi_j)$ 
with some phases $\varphi_j$. It turns out, though, that only $\varphi_j=\pi j$ gives a stable 
mean-field solution.
The Fourier transformed kernel matrix in Eq.~(\ref{eq:MFH}) is a $4\times 4$ matrix
with the two-fold degenerate eigenvalues $\pm E^{}_{\bq}$ with $E^{}_{\bq} = \sqrt{\Delta^2+h^2_{1;\bq}+h^2_{2;\bq}}$,
cf. Appendix~\ref{app:Eigenbasis}. With this result we can treat the grand-canonical free-energy 
\begin{equation}
F = -\beta^{-1}\log{\rm Tr}~e^{-\beta H}
\ , \ \ \beta=1/k_B T
\end{equation}
in mean-field approximation. It is plotted for a subcritical interaction strength $g<g_c$ in the left panel of in Fig.~\ref{fig:Potential} 
and for $g>g_c$ in the right panel. Thus, depending on the interaction strength $g$, the free energy 
has either a single minimum at $\Delta=0$ for interaction $g\le g_c$ 
or a continuously degenerate minimum for $g>g_c$. The continuous degeneracy implies a gapless Goldstone mode.
The corresponding phase transition and its consequences for transport properties and the gapless fluctuations of the order parameter
will be discussed subsequently. 

\begin{figure*}[t]
\includegraphics[width=7cm]{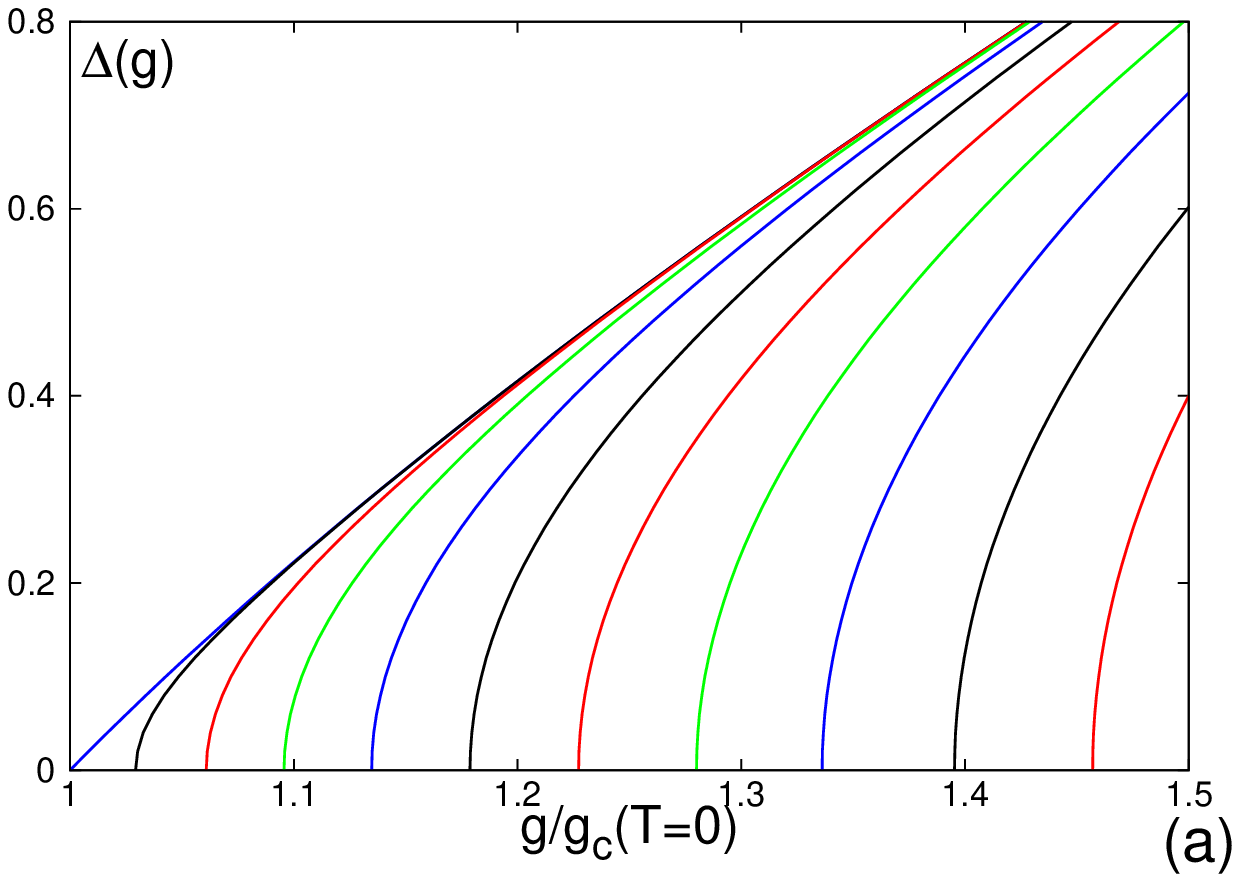}
\includegraphics[width=7cm]{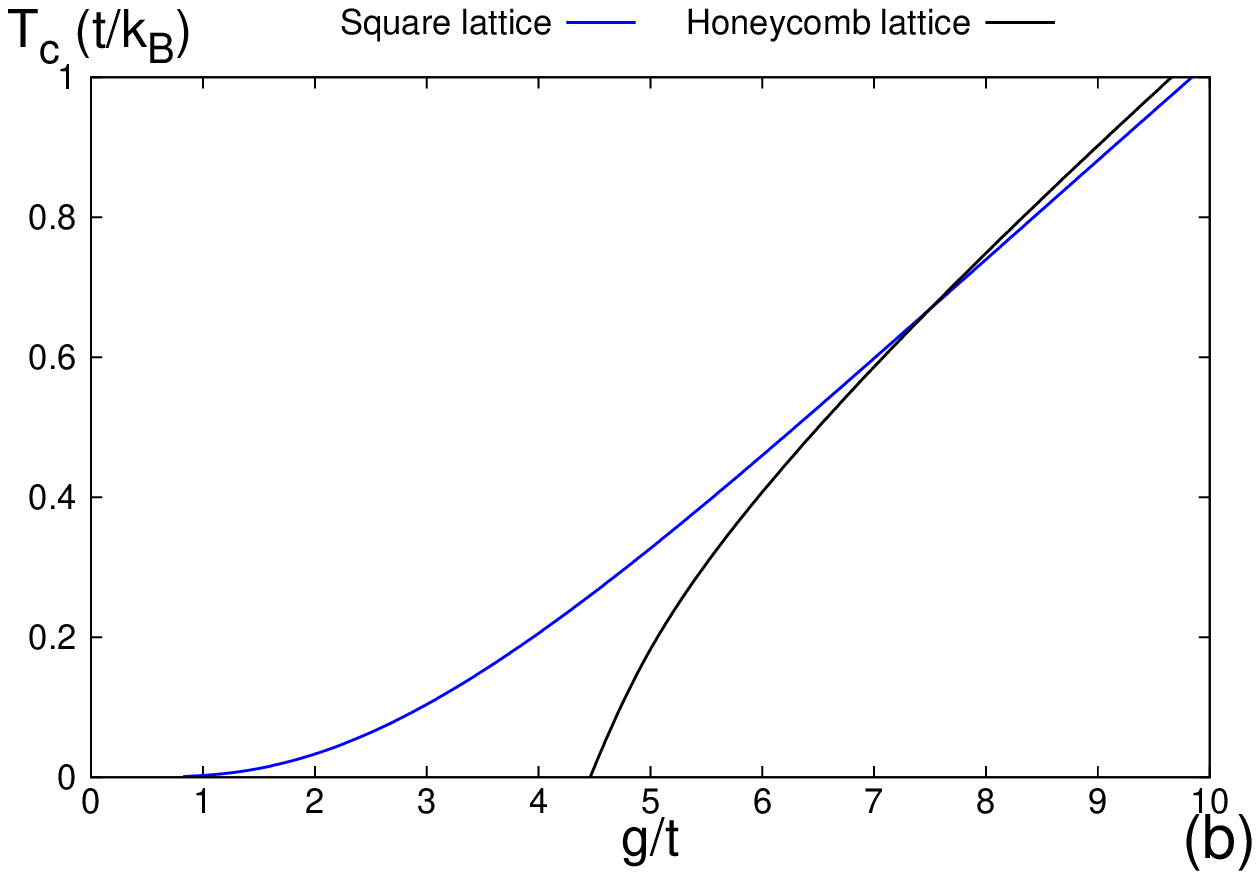}
\caption{
(a): Solution of the gap equation for different temperature values from left to right 
$T/t=0,0.05,0.1,0.15,0.2,0.25,0.3,0.35,0.4,0.45,0.5$.
(b): Comparison of the critical temperatures as functions of the interaction strength 
for the square lattice (blue (gray in printed version) curve) and the honeycomb lattice 
(black curve) in units of the hopping amplitude $t$. 
}
\label{fig:GapEq}
\end{figure*}

To obtain the value of $\Delta$ we solve the mean-field equation $\delta_\Delta F=0$. This reads for solutions $\Delta\ne0$
\begin{equation}
\label{eq:MFEq2}
\frac{1}{g} 
=\frac{1}{4}\int ~ \frac{\tanh(\sqrt{E^2+\Delta^2}/2k^{}_BT)}{\sqrt{E^2+\Delta^2}}\rho(E)dE
\ ,
\end{equation}
where 
$\rho(E)$ is the density of states (DOS). The solution of this equation provides us the value of $\Delta$
as a function of temperature and coupling strength $g$. This is visualized in the left panel of Fig.~\ref{fig:GapEq}.
Moreover, the critical temperature $T_c$ for the pairing transition is plotted as a function of $g$ in the right panel
of Fig.~\ref{fig:GapEq}. Here we compare the electron dispersion of a honeycomb lattice (with a linearly vanishing DOS at $E=0$) 
and of the square lattice (with a logarithmically divergent DOS at $E=0$). The honeycomb lattice has a non-zero threshold of $g$
for a pairing transition, whereas a square lattice has always a non-zero transition temperature for any non-zero interaction.
In contrast to the constant DOS ${\bar \rho}$, which gives the critical temperature $T_c \propto t\exp(-1/g{\bar\rho})$, 
the logarithmic DOS causes a renormalization of the effective coupling $g{\bar\rho}$ to larger values. It is also possible to 
use other dispersions to enhance the critical temperature, e.g., with a van Hove singularity at the Fermi level \cite{berman19}.

\no
\subsection{Duality transformation}
It should be noticed that we can replace the electrons in the lower layer by holes
and the repulsive interaction by an attractive interaction $g\to-g$.
The resulting hopping Hamiltonian $H_\u-H_\d-H_I$ describes
electrons in the upper layer and holes in the lower layer which attract each other through
an attractive interlayer Coulomb interaction. This duality transformation can also be applied to the 
mean-field Hamiltonian $H_{\rm MF}$. It amounts to the replacement of the staggered order
parameter of the electronic system by a uniform order parameter for the electron-hole system.
The latter means formally 
replacing the Pauli matrix $\sigma_3$ by the $2\times2$ unit matrix $\sigma_0$ in Eq. ({\ref{eq:MFH}).
This leads to the following mean-field Hamiltonian of the exciton gas~\cite{lozovik76} 
\begin{equation}
\label{eq:MFH2}
{\rm H}^{}_{\rm exc} = 
\sum_{\bR\bR'}
\left( 
\begin{array}{c}
\hat c^\dag_{\uparrow;\bR} \\ \\
\hat c^\dag_{\downarrow;\bR}
\end{array}
\right)^{\rm T}
\left(
\begin{array}{ccc}
\sum_j h^{}_{j;\bR\bR'}\sigma^{}_j & & \Delta \sigma^{}_0\delta^{}_{\bR\bR'} \\
 \\
 \Delta \sigma^{}_0\delta^{}_{\bR\bR'} & & -\sum_j h^{}_{j;\bR\bR'}\sigma^{}_j
\end{array}
\right)
\left( 
\begin{array}{c}
\hat c^{}_{\uparrow;\bR'} \\ \\
\hat c^{}_{\downarrow;\bR'}
\end{array}
\right),
\end{equation}
where $\hat c^\dagger_{\downarrow;\bR'}$ is the creation operator for a hole (i.e., an annihilation
operator of an electron) in the lower layer.
This Hamiltonian has the same dispersion $E^{}_{\bq}$ as the Hamiltonian $H_{\rm MF}$ of the
electronic double layer, implying that the two problems can be mapped formally onto each other. 
In other words, both systems describe pairing of particles, where
the main difference is that the order parameter is uniform in the electron-hole
double layer and staggered with respect to the two sublattices in the electron-electron double layer.
Here it should be pointed out that a modulated order parameter is a common phenomena in many condensed matter systems,
which can be of different origin. Typical cases are the Fulde-Ferrell-Larkin-Ovchinnikov phase in 
superconductors~\cite{fulde64,larkin65} 
or the formation and melting of the Wigner crystal phase in bilayer systems, which has been intensively studied in 
Refs.~[\onlinecite{Neilson1991,Peeters1996,Peeters1997,Peeters1999}].

\begin{figure*}[t]
\includegraphics[width=7.5cm]{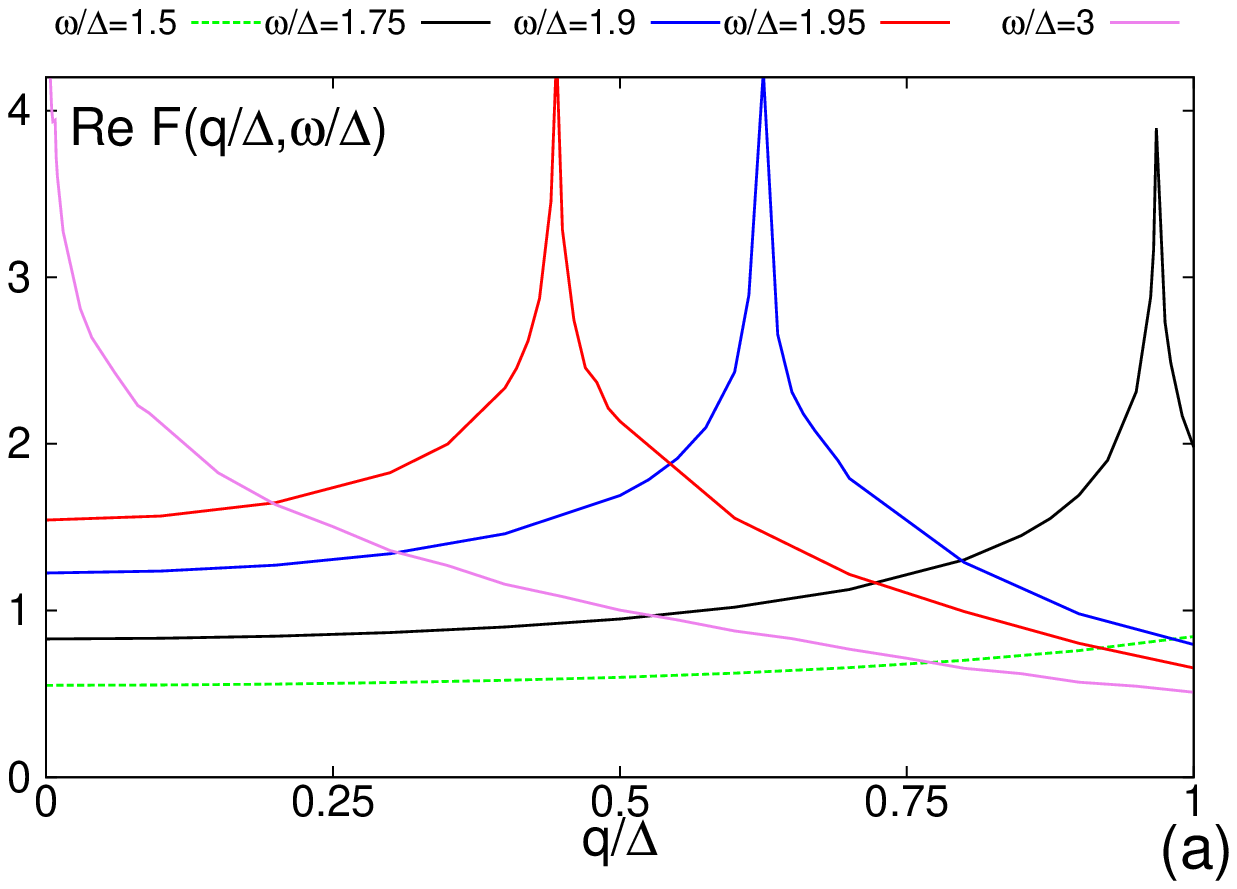}
\includegraphics[width=7.5cm]{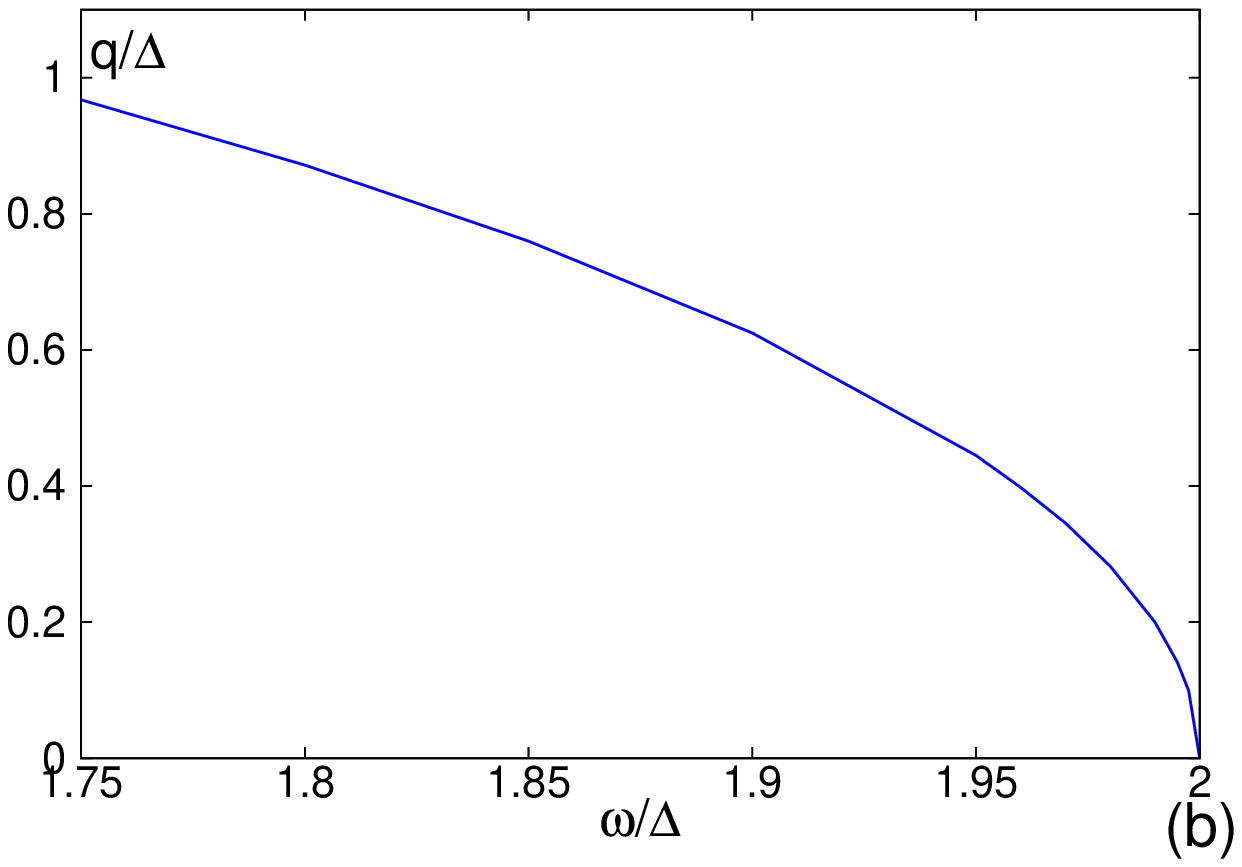}
\caption{
(a): The real part of the current-current correlator of Eq.~(\ref{eq:ScalFunc}) as function of the momentum for different frequencies.
(b): Position of the singularity in the correlator at the edge of the gap for different frequencies and momenta
according to Eq. (\ref{sing}). 
}
\label{fig:Correlator}
\end{figure*}

\no
\subsection{Current-current correlations}
To analyze the effect of the interlayer Coulomb interaction we create a local current density
at time $t=0$ at site $\bR'$ on the lower layer which oscillates with frequency $\omega$. Then 
we measure the local current at time $t>0$ at site $\bR$ on the upper layer, in analogy to the
drag effect~\cite{kulakovskii04,lozovik76} . This measurement
probes how much a local current in the lower layer will generate currents in the
upper layer. It can be described in terms of linear response theory, which is associated with
the current-current correlator between the two layers and reads
\begin{equation}
\label{corr1}
C^{}_{\mu\nu}(\bR,\bR';\omega) =i\lim_{\alpha\to0}\intop^{\infty}_{0} dt~e^{-i(\omega-i\alpha)t}
\langle\hat j^{}_{\mu\uparrow}(\bR,t)\hat j^{}_{\nu\downarrow}(\bR',0)\rangle, \;\;\; 
\langle \cdots \rangle = {\rm Tr}\left[e^{-\beta H}\cdots \right]/{\rm Tr}e^{-\beta H},
\end{equation}
with a positive parameter $\alpha$. For low temperatures the Fourier components 
of the correlator become, cf. Appendix~\ref{app:cc-corr}:
\begin{equation}
\label{corr2}
{\tilde C}^{}_{\mu\mu}(\bq;\omega) \approx \frac{\Delta}{4\pi} F\left(\frac{q}{\Delta};\frac{\omega}{\Delta}\right)
\end{equation}
with
\begin{equation}
\label{eq:ScalFunc}
F\left(\bar{q};\bar{\omega}\right) = \intop^\infty_1 dx\left[
\frac{1}{\sqrt{x^2(-i\alpha+\bar\omega+2x)^2-\bar q^2(x^2-1)}} -
\frac{1}{\sqrt{x^2(-i\alpha+\bar\omega-2x)^2-\bar q^2(x^2-1)}} \right].
\end{equation}
Its real part is plotted in Fig.~\ref{fig:Correlator} and exhibits a singularity at $q_{\rm sing}(\omega)$ with
\begin{equation}
q^{}_{\rm sing}(\omega) = \sqrt{4\Delta^2-\omega^2} .
\label{sing}
\end{equation}
The singular wavevector is visualized in the right panel of Fig.~\ref{fig:Correlator}.
It corresponds with an inverse length scale (wavelength) for the spatial distribution 
of the current in the upper layer, caused by the local current in the lower layer with frequency $\omega$. 
As long as the frequency is less than the gap $2\Delta$, there is a mode with a finite wavelength. When
$\omega$ is equal to the gap, this wavelength diverges and the strongest correlation appears with a zero
mode $q^{}_{\rm sing}(2\Delta)=0$. And finally, when $\omega$ exceeds the gap, $q^{}_{\rm sing}(\omega)$ becomes
imaginary, which indicates an exponential spatial decay of the strongest correlation.  

The $\omega$ dependence of the $q=0$ component of the current-current correlator also has a charcteristic behavior at the gap:
\begin{eqnarray}
{\rm Re}~{\tilde C}^{}_{\mu\mu}(0;\omega)  &=& \frac{1}{4\pi}\frac{\Delta^2}{\omega}\log\left|\frac{2\Delta+\omega}{2\Delta-\omega}\right|,\\
{\rm Im}~{\tilde C}^{}_{\mu\mu}(0;\omega)  &=& \frac{1}{4}\frac{\Delta^2}{\omega}\left[\Theta(\omega-2\Delta) + \Theta(-\omega-2\Delta)\right].
\end{eqnarray}

\begin{figure*}[t]
\includegraphics[width=7cm]{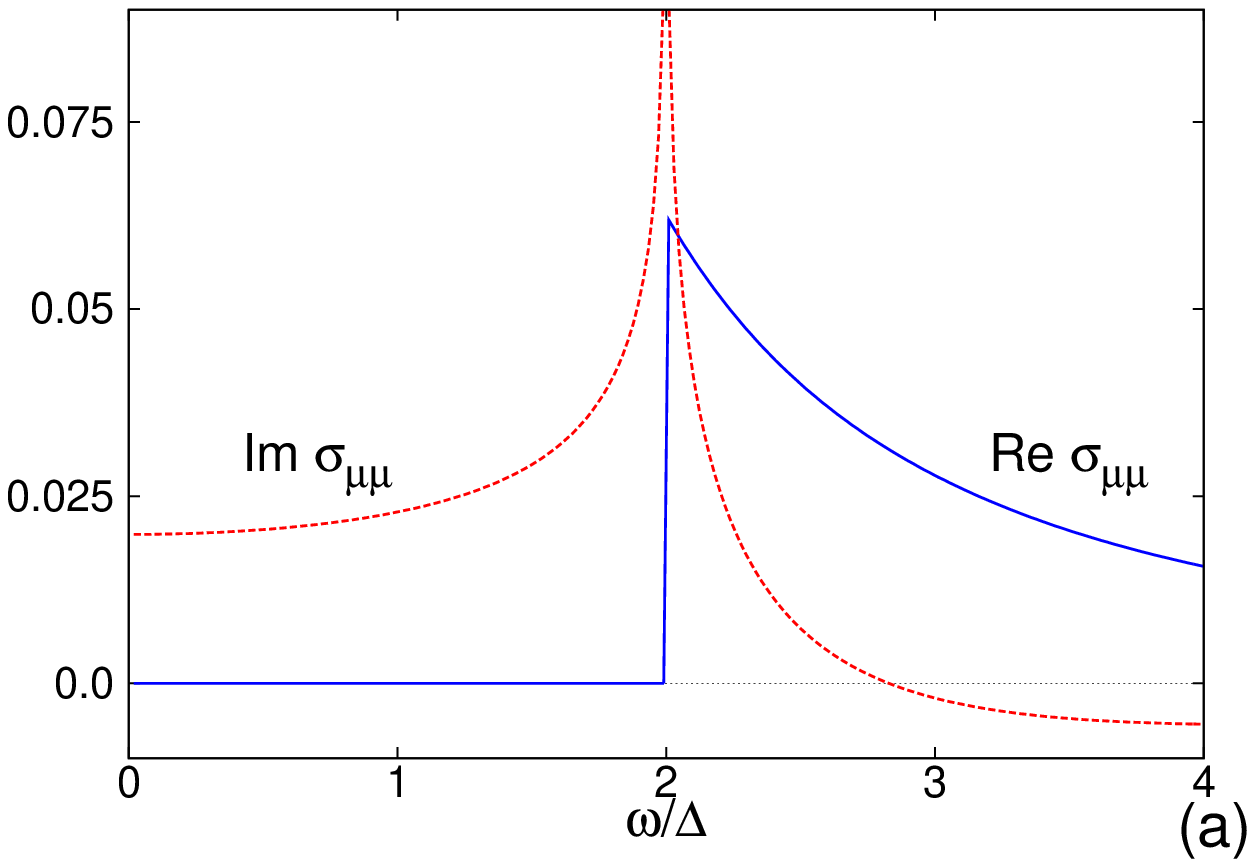}
\includegraphics[width=7cm]{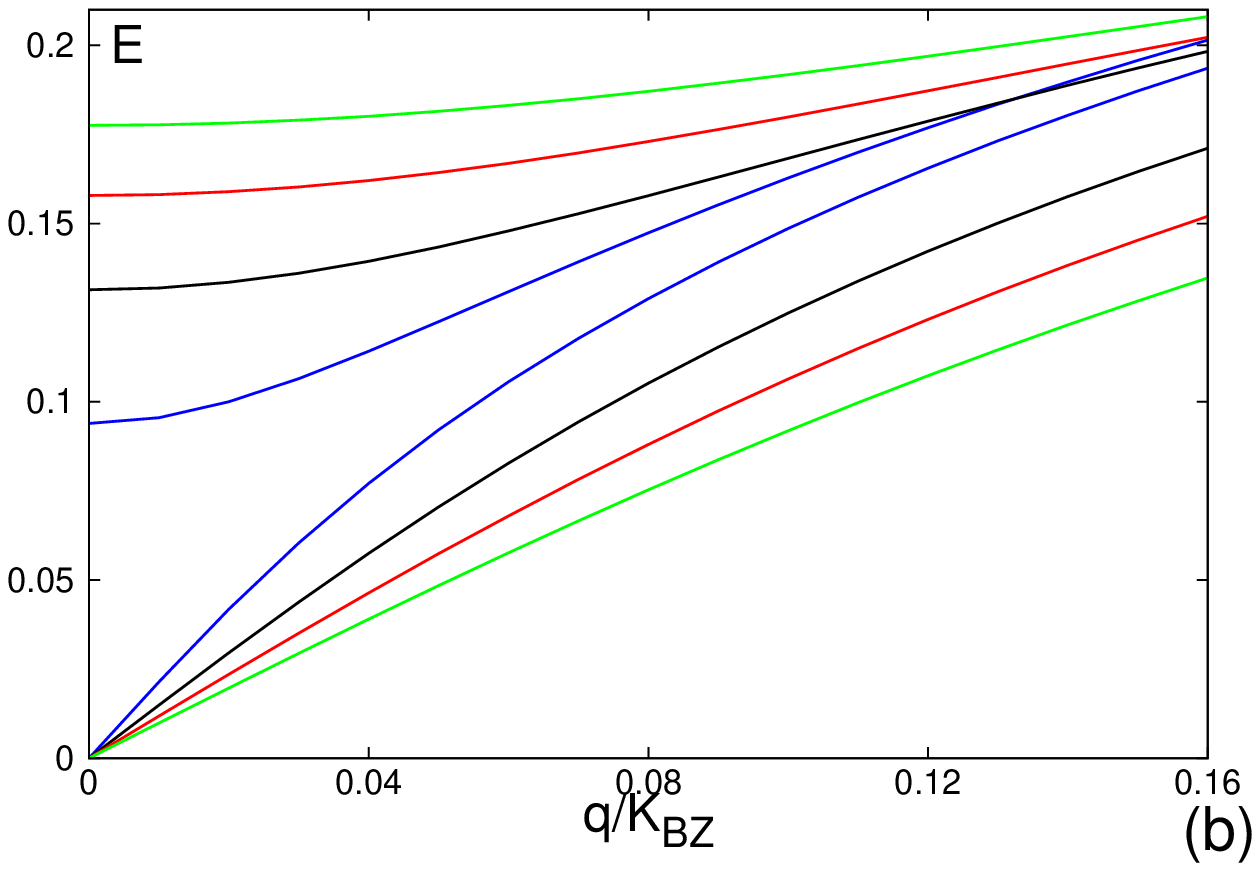}
\caption{
(a): Real and imaginary part of the conductivity according to equations (\ref{eq:RePart}) and (\ref{eq:ImPart}).
(b): Gapless and gapped spectral branches of the order parameter fluctuations, calculated 
for $\Delta=0.125t$ (blue lines), $\Delta=0.25t$ (black lines), $\Delta=0.375t$ (red lines) and 
$\Delta=0.5t$ (green lines). Only positive energies at positive momenta are shown. The momentum 
$\rm K^{}_{BZ}$ is the square root of Brillouin zone area.
(For paper printed black-white version: smallest $\Delta$ has been used to calculate the
lowest gapped and upper higher gapless curve.)
}
\label{fig:ACCond}
\end{figure*}

\no
\subsection{ Interlayer conductivity}
From the current-current correlator we obtain the inter-layer conductivity tensor as \cite{allen}
\begin{equation}
\sigma^{}_{\mu\nu}(\omega,T) \sim \frac{i}{\omega}\left[{\tilde C}^{}_{\mu\nu}(0;\omega)-{\tilde C}^{}_{\mu\nu}(0;0)\right].
\end{equation}
This gives for the diagonal conductivity 
\begin{equation}
\sigma^{}_{\mu\mu}(\omega,T) =  i\int\frac{d^2k}{(2\pi)^2} 
\frac{f{}_\beta(-E^{}_\bk) - f^{}_\beta(E^{}_\bk)}{4E^3_\bk}
\left[\frac{\Delta^2}{\omega-i\alpha + 2E^{}_\bk} - \frac{\Delta^2}{\omega-i\alpha - 2E^{}_\bk} \right]
\end{equation}

\no
with the Fermi-Dirac function $f^{}_\beta(E^{}_\bk)$ at inverse temperature $\beta$.
The real and imaginary parts read at low temperatures 
\begin{equation}
\label{eq:RePart}
{\rm Re}~\sigma^{}_{\mu\mu}(\omega,0)=
\frac{\Delta^2}{4\omega^2}\left[\Theta\left(\omega - 2|\Delta|\right) - \Theta\left(-\omega - 2|
\Delta|\right)\right]
\end{equation}
\begin{eqnarray}
\label{eq:ImPart}
{\rm Im} ~\sigma^{}_{\mu\mu}(\omega,0) 
=\frac{1}{4\pi}\frac{\Delta^2}{\omega^2}\log\left|\frac{4|\Delta|^2}{\omega^2-4|\Delta|^2}\right| 
.
\end{eqnarray}
The real part of the conductivity is zero when the frequency is less than the gap. But when $\omega$ reaches the gap,
it jumps to a finite value, indicating that there is a current between the two layers when the photon energy $\hbar\omega$ 
is equal or larger than the gap. This jump is similar to the jump observed for pairing in electron-hole layers at $\omega=0$
and $T>0$ \cite{vignale96}.
The imaginary part, on the other hand, is always nonzero in the pairing phase and has a
logarithmic divergence when the real part jumps (cf. Fig. \ref{fig:ACCond}).

\vspace{1mm}
\no
\subsection{Quantum fluctuations of the order parameter} 
Quantum fluctuations around the mean-field solution consist of
two spectral branches in the low-energy sector: a gapless branch due the degenerate ring of Fig. \ref{fig:Potential} and a 
gapped branch. The dispersion of both branches can be evaluated within the low-energy approximation and a gradient expansion $p\sim 0$, 
cf. Appendix~\ref{app:Fluctuations}:
\begin{equation}
E^{}_{\rm long}(p)\approx \frac{p^2+6\Delta^2}{12\pi\Delta}\ , \;\;\;
E^{}_{\rm trans}(p)\approx \frac{p^2}{8\pi\Delta}\ ,
\label{eq:Bogolubov}
\end{equation}
indicating a stable mean-field solution.
The result for a honeycomb lattice is visualized in the right panel of Fig.~\ref{fig:ACCond}.

\vspace{1mm}
\no
\subsection{Discussion and summary}
In an electronic double layer there is an interlayer pairing transition due to a strong Coulomb interaction.
Its critical temperature $T_c$ increases almost linearly with large coupling strength (cf. Fig. \ref{fig:GapEq}).
For weak coupling the behavior depends on the details of the DOS: On the honeycomb lattice $T_c=0$ if $g<g_c$
and for the square lattice $T_c$ vanishes exponentially with $1/g$.
The pairing phase is accompanied by strong interlayer current-current correlations. These correlations diverge
if the relation $q^2+\omega^2=\Delta^2$ is satisfied (cf. Fig. \ref{fig:Correlator}). This
reflects a long-range current-current correlation if $\omega^2=\Delta^2$ holds. The real part
of the interlayer conductivity is non-zero only for frequencies $\omega$ larger than the gap of the pairing phase.
Although there is a divergent current-current correlation for $\omega$ less than the gap, the real part of 
the interlayer conductivity $\sigma_{\mu\mu}$ is zero.
Due to this characteristic behavior the drag effect would be a good candidate to identify experimentally the pairing 
phase in electronic double layers. The effective coupling strength can be tuned by the interlayer distance and by 
the coefficient of the interlayer dielectric material, which enables the experiment to drive through the pairing 
transition. Moreover, the lattice structure of the layers determines the density of states. This could be used,
for instance by applying a strain field, to change the pairing transition temperature. And finally, the duality
relation provides a basis to compare the large number of experiments on excitonic double layers with more recent
experiments on electronic double layers in order to clarify some of the recent experimental results.

\vskip0.1cm
\no
\section*{Acknowledgments:} 
This research was supported by a grant of the Julian Schwinger Foundation for Physics Research.
A.S. expresses his gratitude to the MPI-PKS Dresden for the support through the Visitors Program.  
Yu.E.L. was supported through the grants RFBR 20-02-00410 and 20-52-00035.

\appendix

\section{Effective electron-electron interaction}
\label{app:EffInt}

Using the results from Ref.~[28], 
the effective electron-electron interaction parameter $g$ is given by
\begin{eqnarray}  \label{ehint}
\frac{2}{g} = \frac{1}{|B|} \int_{B} \frac{1}{U_\mathbf{p}}d^{2}p \ ,
\end{eqnarray}
where $B$ is the Brillouin zone, $|B|$ its area and
\begin{equation}
U_{\mathbf{p}}\approx \frac{\hbar \bar{U}e^{-pD/\hbar }}{p+a(p)}
\label{static_screening}
\end{equation}
is the screened Coulomb interaction.
${\bar{U}}=2\pi |B| \kappa e^{2}/(\hbar^{2}\varepsilon)= 4\pi^{2}
\kappa e^{2}/(\varepsilon b^{2})$ is the interaction strength, $\varepsilon$
is the dielectric constant $\kappa =9\times 10^{9} \ \mathrm{N \times m^{2}/C^{2}}$, $e$ is the
electron charge, $m$ the electron mass, $a=\hbar ^{2}\varepsilon /(\kappa e^{2} m)$,
and $D$ is the thickness of the dielectric interlayer, which fills the space between the two layers. 
$a(p)$ is the material specific function
\begin{eqnarray}  \label{LY}
a(p) = \frac{4\hbar}{a} +\frac{4\hbar^{2}\left(
1-\exp \left( -2pD/\hbar\right) \right)}{a^2 p} \ .
\end{eqnarray}

\section{Mean-field ansatz}
\label{app:MeanField}

Using local canonical anticommutation relations for fermionic second quantization operators
\begin{equation}
\{\hat c^\dag_{i,s},\hat c^{}_{i',s'}\} = \delta^{}_{ii'}\delta^{}_{ss'},
\end{equation}
where $s,s'$ refer to the layer index and $\hat c^{}_{1,s}= c^{}_{s}$, $\hat c^{}_{2,s}= d^{}_{s}$ and correspondingly 
for $\hat c^\dag_{i,s}$ as defined in the main text, 
it is easy to show the following identity which is valid at every lattice site $\bR$ (from right to left):
\begin{equation}
\label{eq:Prep}
\frac{g}{2}n^{}_\uparrow n^{}_\downarrow = \frac{g}{2}(n^{}_\uparrow + n^{}_\downarrow) - 
\frac{g}{8}\sum^3_{\mu=0}\left[ 
(\hat c^\dag_\uparrow\sigma^{}_\mu\hat c^{}_\downarrow)(\hat c^\dag_\downarrow\sigma^{}_\mu\hat c^{}_\uparrow) + 
(\hat c^\dag_\downarrow\sigma^{}_\mu\hat c^{}_\uparrow)(\hat c^\dag_\uparrow\sigma^{}_\mu\hat c^{}_\downarrow)
\right],
\end{equation}
where $\sigma^{}_{\mu=1,2,3}$ are Pauli matrices and $\sigma^{}_0$ is a two dimensional unity matrix. In mean-field 
approximation we put densities in both layers on the same value (zero at half filling) and introduce the 
order parameters
\begin{equation}
\Delta^{}_\mu = -\frac{g}{4} \langle \hat c^\dag_{\uparrow;\bR}\sigma^{}_\mu\hat c^{}_{\downarrow;\bR} \rangle , \;\;\;  
\Delta^\ast_\mu = -\frac{g}{4} \langle 
\hat c^\dag_{\downarrow;\bR}\sigma^{}_\mu\hat c^{}_{\uparrow;\bR} \rangle. 
\end{equation}
The only order parameter which has a stable mean-field solution is $\Delta^{}_3$ which is exclusively considered in this paper. Because of the $U(1)$ symmetry
of the problem there are infinitely many solutions which describe thermodynamically one and the same system. In the special solution which is discussed in the 
main text we consider a purely real order parameter and call it $\Delta$. Eq.~(\ref{eq:Prep}) is easily generalized to generic non-local 
interactions on bipartite lattices 
\begin{equation}
\sum_{\bR,\bR'} U^{}_{s\bR;s'\bR'} n^{}_{s;\bR}n^{}_{s';\bR'}, \;\;\; n^{}_{s;\bR}= (\hat c^\dag_{s;\bR}\hat c^{}_{s;\bR}) 
= c^\dag_{s;\bR}c^{}_{s;\bR} + d^\dag_{s;\bR}d^{}_{s;\bR}.
\end{equation}
Here we can prepare the product of two densities as
\begin{equation}
n^{}_{s;\bR}n^{}_{s';\bR'} =  n^{}_{s;\bR} + n^{}_{s';\bR'} - \frac{1}{4}\sum^3_{\mu=0}\left[  
(\hat c^\dag_{s;\bR}\sigma^{}_\mu\hat c^{}_{s';\bR'})(\hat c^\dag_{s';\bR'}\sigma^{}_\mu\hat c^{}_{s;\bR}) + 
(\hat c^\dag_{s';\bR'}\sigma^{}_\mu\hat c^{}_{s;\bR})(\hat c^\dag_{s;\bR}\sigma^{}_\mu\hat c^{}_{s';\bR'})
\right],
\end{equation}
which is valid for all combinations of indices, i.e. for $s=s'$ and $\bR=\bR'$ too.

\section{Eigenbasis of the mean-field Hamiltonian}
\label{app:Eigenbasis}

In general the Fourier transformed kernel matrix of the mean-field Hamiltonian Eq.~(\ref{eq:MFH}) 
\begin{equation}
{\cal K}^{}_\bq = 
\left(
\begin{array}{ccc}
 \sum_j h^{}_{j;\bq}\sigma^{}_j & & \Delta e^{-i\phi}\sigma^{}_3 \\
 \\
 \Delta e^{i\phi}\sigma^{}_3 & & \sum_j h^{}_{j;\bq}\sigma^{}_j
\end{array}
\right).
\end{equation}
It has folowing  orthogonalized and normalized eigenvectors:

\vspace{2mm}
1) For the negative eigenvalue (lower band) $-E^{}_\bq=-\sqrt{h^2_{1,\bq}+h^2_{2,\bq}+\Delta^2}$:
\begin{equation}
\label{eq:PEV}
|-,1,\bq\rangle = \frac{1}{\sqrt{2}E^{}_\bq} 
\left( 
\begin{array}{c}
-e^{-i\phi}(h^{}_{1,\bq}-i h^{}_{2,\bq}) \\
e^{-i\phi}E^{}_\bq \\
0 \\
\Delta
\end{array}
\right), \;\;\; 
|-,2,\bq\rangle  = \frac{1}{\sqrt{2}E^{}_\bq} 
\left( 
\begin{array}{c}
-e^{-i\phi} \Delta\\
0 \\
E^{}_\bq\\
-(h^{}_{1,\bq}+i h^{}_{2,\bq})
\end{array}
\right), 
\end{equation}

2) For the positive eigenvalue (upper band) $+E^{}_\bq=+\sqrt{h^2_{1,\bq}+h^2_{2,\bq}+\Delta^2}$:
\begin{equation}
\label{eq:NEV}
|+,1,\bq\rangle  = \frac{1}{\sqrt{2}E^{}_\bq} 
\left( 
\begin{array}{c}
-e^{-i\phi}(h^{}_{1,\bq}-i h^{}_{2,\bq}) \\
-e^{-i\phi}E^{}_\bq \\
0 \\
\Delta
\end{array}
\right), \;\;\; 
|+,2,\bq\rangle  = \frac{1}{\sqrt{2}E^{}_\bq} 
\left( 
\begin{array}{c}
e^{-i\phi} \Delta\\
0 \\
E^{}_\bq\\
h^{}_{1,\bq}+i h^{}_{2,\bq}
\end{array}
\right). 
\end{equation}

\section{Current-current correlations}
\label{app:cc-corr}

We expand the correlator of Eq.~(\ref{corr1}) in terms of the eigenstates $\{|n\rangle\}$ 
with the corresponding eigenvalues $\{E_n\}$ of the mean-field Hamiltonian 
Eq.~(\ref{eq:MFH}), cf. Appendix~\ref{app:Eigenbasis}. This gives
\begin{eqnarray}
C^{}_{\mu\nu}(\bR,\bR';\omega)   &=& \lim_{\alpha\to0}\sum_{n,m}\left[f{}_\beta(E^{}_m) - f^{}_\beta(E^{}_n)\right]~\frac{\langle n| 
\hat j^{}_{\mu \uparrow}(\bR) |m \rangle \langle m| \hat j^{}_{\nu \downarrow}(\bR') |n \rangle}{\omega-i\alpha-E^{}_m+E^{}_n}
\ ,
\end{eqnarray}
where  $f^{}_\beta(E^{}_n)$ is the Fermi-Dirac distribution at inverse temperature $\beta$.
Then the Fourier transformed correlator becomes
\begin{equation}
{\tilde C}^{}_{\mu\nu}(\bq;\omega)=\int\frac{d^2k}{(2\pi)^2}
\sum_{n,m}\left[f{}_\beta(E^{}_{m,\bk}) - f^{}_\beta(E^{}_{n,\bk})\right]~
\frac{\langle n,\bk+\bq| \hat j^{}_{\mu \uparrow} |m,\bk \rangle \langle m,\bk| \hat j^{}_{\nu \downarrow} 
|n,\bk+\bq \rangle}{\omega-i\alpha-E^{}_{m,\bk}+E^{}_{n,\bk+\bq}}
\ ,
\end{equation}
which can be expanded for small $\bq$ to give
\begin{equation}
{\tilde C}^{}_{\mu\mu}(\bq;\omega) \approx \int\frac{d^2k}{(2\pi)^2} 
\frac{f{}_\beta(-E^{}_\bk) - f^{}_\beta(E^{}_\bk)}{2E^2_\bk}\left[\frac{\Delta^2}{\omega-i\alpha + 2E^{}_\bk +
\frac{\bq\cdot\bk}{E^{}_\bk}} - \frac{\Delta^2}{\omega-i\alpha - 2E^{}_\bk -  \frac{\bq\cdot\bk}{E^{}_\bk}}\right].
\end{equation}
This correlation is non-zero only in the gapped phase, while it vanishes in the gapless phase.
The angular integration is performed using the relation
\[
\frac{1}{2\pi} \int^{2\pi}_0\frac{~d\phi}{A\pm B\cos\phi} = \frac{1}{\sqrt{(A-B)(A+B)}}
\]
to yield at low temperatures Eq.~(\ref{corr2}).

\section{Quantum fluctuation term}
\label{app:Fluctuations}

To obtain the spectra of order parameter fluctuations we assume small quantum deviations from the mean-field value $\Delta^\ast\to\Delta^\ast\delta^{}_{\bR\bR'} 
+ B^\dag_{\bR\bR'}$ and $\Delta^{}\to\Delta^{}\delta^{}_{\bR\bR'} + B^{}_{\bR\bR'}$. The effective Hamiltonian for the fluctuating order parameter $B$ can be 
found from the second order of the Schr\"odinger perturbation theory if we treat the term 
\begin{equation}
{\cal K}^{(1)}_{\bR\bR'} = 
\left(
\begin{array}{ccc}
 0 & & B^\dag_{\bR\bR'}\sigma^{}_3 \\\\
 B^{}_{\bR\bR'}\sigma^{}_3 & & 0
\end{array}
\right), \;\; {\rm i.e.} \;\;
{\cal K}^{(1)}_{\bp} = 
\left(
\begin{array}{ccc}
 0 & & B^\dag_{\bp}\sigma^{}_3 \\\\
 B^{}_{-\bp}\sigma^{}_3 & & 0
\end{array}
\right)
\end{equation}
as a small perturbation. (The first order term is zero because of the mean-field condition.) In Fourier representation we have
\begin{equation}
\delta H = \frac{1}{2}\int\frac{d^2p}{|BZ|} \int\frac{d^2q}{|BZ|} \sum_{n,m}\left[f^{}_\beta(E_{n,\bq})-f^{}_\beta(E_{m,\bq})\right]
\frac{\langle n,\bq+\bp|{\cal K}^{(1)}_{\bp}|m,\bq\rangle\langle m,\bq|{\cal K}^{(1)}_{-\bp}|n,\bq+\bp\rangle}{E_{n,\bq+\bp}-E_{m,\bq}-i\alpha} , 
\end{equation}
$\alpha$ is to be send to zero and the average is performed over the eigenstates of the mean-field Hamiltonian given in the Appendix~\ref{app:Eigenbasis}. 
Because of the Fermi functions combination only interband matrix elements contribute. After some algebra we get to 
\begin{equation}
\label{eq:FLC}
\delta H = \int\frac{d^2p}{|BZ|} 
\left(
\begin{array}{c}
B^\dag_\bp \;\; B^{}_{-\bp} 
\end{array}
\right)
\left(
\begin{array}{ccc}
 A^{}_{11,\bp} & & A^{}_{12,\bp} \\
 \\
 A^{}_{21,\bp} & & A^{}_{22,\bp} 
\end{array}
\right)
\left(
\begin{array}{c}
B^{}_{\bp} \\\\
B^\dag_{-\bp}
\end{array}
\right),
\end{equation}
where 
\begin{eqnarray}
A^{}_{11,\bp} = A^{}_{22,\bp}   &=& ~\frac{1}{2}\int\frac{d^2q}{|BZ|}~\left[f^{}_\beta(-E_{\bq})-f^{}_\beta(E_{\bq})\right]\frac{h^{}_{\bq}\cdot h^{}_{\bq+\bp}+E^{}_\bq E^{}_{\bq+\bp}}{(E^{}_\bq+E^{}_{\bq+\bp})E^{}_\bq E^{}_{\bq+\bp}}, \\
A^{}_{12,\bp} = A^\ast_{21,\bp} &=& -\frac{1}{2}\int\frac{d^2q}{|BZ|}~\left[f^{}_\beta(-E_{\bq})-f^{}_\beta(E_{\bq})\right]\frac{e^{2i\phi}\Delta^2}{(E^{}_\bq+E^{}_{\bq+\bp})E^{}_\bq E^{}_{\bq+\bp}}, \;\;
\end{eqnarray}
where $h^{}_{\bq}\cdot h^{}_{\bq+\bp}=h^{}_{1,\bq}h^{}_{1,\bq+\bp}+h^{}_{2,\bq}h^{}_{2,\bq+\bp}$. In order to approach the structure 
of elementary excitations above the mean-field ground state we have to subtract Eq.~(\ref{eq:FLC}) from the mean-field background 
energy 
\begin{equation}
\label{eq:FLCmf}
\delta H^{}_{MF} = 
\int\frac{d^2q}{|BZ|} 
~\frac{f^{}_\beta(-E_{\bq})-f^{}_\beta(E_{\bq})}{2E^{}_\bq}
\int\frac{d^2p}{|BZ|} 
~
\left( B^\dag_\bp B^{}_{\bp} 
+
B^{}_{-\bp}B^\dag_{-\bp}\right)
.
\end{equation}
Elementary manipulations turn the combination of Fermi functions into the hyperbolic tangent of Eq.~(\ref{eq:MFEq2}).
We proceed further for the $T=0$ and effective low-energy Dirac approximation. Performing the gradient expansion to quadratic order we get 
\begin{equation}
\delta H^{}_{MF} - \delta H\approx \frac{1}{4\pi}
\int\frac{d^2p}{(2\pi)^2} 
\left(
\begin{array}{c}
B^\dag_\bp \\\\ B^{}_{-\bp} 
\end{array}
\right)^{\rm T}
\left(
\begin{array}{ccc}
\displaystyle \Delta + \frac{5}{12\Delta}p^2 & &\displaystyle e^{-2i\phi}\left(\Delta - \frac{1}{12\Delta}p^2 \right) \\ 
 \\ 
\displaystyle e^{2i\phi}\left(\Delta - \frac{1}{12\Delta}p^2 \right) & & \displaystyle \Delta + \frac{5}{12\Delta}p^2
\end{array}
\right)
\left(
\begin{array}{c}
B^{}_{\bp} \\\\ B^\dag_{-\bp}
\end{array}
\right).
\end{equation}
The diagonalization can be readily done by a unitary rotation into the real base. For small momenta, 
the two (longitudinal and transversal) spectral branches are
\begin{equation}
E^{}_{\rm long}(p)\approx \frac{p^2+6\Delta^2}{12\pi\Delta} , \;\;\;
E^{}_{\rm trans}(p)\approx \frac{p^2}{8\pi\Delta} .
\end{equation}

\end{document}